\documentclass[12pt,thmsa]{article}
\usepackage{amssymb}
\usepackage{sw20lart}
\input{tcilatex}
\begin{document}
\noindent UC-ph/99-50

\author{A. Benslama and N. Mebarki \\
D\'{e}partement de Physique Th\'{e}orique.\\
Universit\'{e} Mentouri. Constantine.\\
Algeria.}
\title{\textbf{Left-Right Gauge Model in Nonassociative Geometry}}
\date{May 1999}
\maketitle

\begin{abstract}
We reformulate the left-right gauge model of Pati-Mohapatra using
nonassociative geometry approach. At the tree level we obtain the mass
relations $M_{W}=\frac{1}{2}m_{t},$ $M_{H}=\frac{3}{2}m_{t}$ and the mixing
angles are $\sin ^{2}\theta _{W}=\frac{3}{8}$ and $\sin ^{2}\theta _{s}=%
\frac{3}{5},$ which are identical to the ones obtained in $SO(10)$ GUT
.\newpage
\end{abstract}

\section{Introduction}

One of the greatest achievement of noncommutative geometry (abbreviated 
\textbf{NCG} hereafter) is its geometrization of the standard model (\cite
{Connes1},\cite{Connes2},\cite{Model}). NCG provides a framework where the
Higgs boson $H$ may be introduced on the same level as $W^{\pm }$ and $Z$
bosons. In this approach we introduce additional discrete dimensions to the
four-dimensional space-time. If the gauge bosons are associated to the
continuous directions, the Higgs boson results from gauging the discrete
directions. However one of the shortcomings of NCG is its non accessibility
to grand unified theories GUT (\cite{Mangano},\cite{Brout}). The other
formulation of NCG due to Coquereaux \cite{Coquereaux} does not suffer from
this problem (\cite{Okum1}-\cite{Konisi}). There are some variant of Connes'
theory (\cite{Chams1},\cite{Chams2}) where we use an auxiliary Hilbert space
to fit GUT.

R. Wulkenhaar has recently succeeded in formulating another type of geometry
which shares a lot of common points with NCG \`{a} la Connes \cite
{Wulkenhaar}. The theory was baptized ''nonassociative geometry'' \textbf{NAG%
}. The main difference with the two theories is that NAG is based on unitary
Lie algebra instead of unital associative $\star $-algebra in the case of
NCG. Its application to several physical models has been successful (\cite
{Wulkenhaar1}-\cite{Wulk1}). 

We try using this approach to reformulate the left-right model LRM  (\cite
{Mohapatra},\cite{moh}). One of the features of NAG (and NCG) is its
geometric explanation of the spontaneous breakdown of gauge symmetry. Our
aim is to investigate whether NAG could encompass also the parity violation.

In section 2 we present the main elements to formulate gauge theories using
NAG. In section 3 we derive the bosonic action of left-right model and
finally we discuss our results.

\section{Some elements of Nonassociative Geometry}

NAG is based on the L-cycle ($\frak{g,}\mathcal{H},D,\pi ,\Gamma )$ $\cite
{Wulkenhaar}$%
\begin{equation}
\frak{g=}C^{\infty }(X)\otimes \frak{a}
\end{equation}
where $C^{\infty }(X)$ is the algebra of smooth functions on the manifold X
and $\frak{a}$ is the matrix Lie algebra. $\mathcal{H}$ is the Hilbert space 
\begin{equation}
\mathcal{H=}L^{2}(X,S)\otimes \Bbb{C}^{F}
\end{equation}
where $L^{2}(X,S)$ is the Hilbert space of spinors.

$D$ is the total Dirac operator given by: 
\begin{equation}
D=\mathbf{D}\otimes \mathbf{1}_{F}+\gamma ^{5}\otimes \mathcal{M}
\end{equation}
where $\mathbf{D}$ is the Dirac operator associated with the continuous
algebra $C^{\infty }(X)$ and $\mathcal{M}$ is associated to the discrete
algebra $\frak{a.}$

The representation of $\frak{g}$ on $\mathcal{H}$ is given by 
\begin{equation}
\pi =\mathbf{1\otimes }\hat{\pi},
\end{equation}
where $\hat{\pi}$ is the representation of $\frak{a}$ on $\Bbb{C}^{F}.$

Finally, the graded operator $\Gamma $, acting on $\mathcal{H}$, is given
by: 
\begin{equation}
\Gamma =\gamma ^{5}\otimes \hat{\Gamma},
\end{equation}
where $\hat{\Gamma}$ is the grading operator acting on $\Bbb{C}^{F}.$

The space $\hat{\Omega}^{1}\frak{a}$ is generated by the elements of the
type 
\begin{equation}
\omega ^{1}=\stackunder{z}{\sum }[a^{z},...[a^{1},da^{0}]...]\text{ \qquad }%
a^{i}\in \frak{a.}
\end{equation}

The representation $\hat{\pi}$ acts on the space $\hat{\Omega}^{1}\frak{a}$
as: 
\[
\hat{\pi}:\hat{\Omega}^{1}\frak{a\longrightarrow }\Bbb{M}_{F}(\Bbb{C)} 
\]

\begin{equation}
\tau ^{1}=\hat{\pi}(\omega ^{1}):=\stackunder{z}{\sum }[\hat{\pi}(a^{z}),...[%
\hat{\pi}(a^{1}),[-i\mathcal{M},\hat{\pi}(a^{0})]]...]\text{ }
\end{equation}

We define also the mapping: 
\[
\hat{\sigma}:\hat{\Omega}^{1}\frak{a\longrightarrow }\Bbb{M}_{F}(\Bbb{C)} 
\]

\begin{equation}
\hat{\sigma}(\omega ^{1}):=\stackunder{z}{\sum }[\hat{\pi}(a^{z}),...[\hat{%
\pi}(a^{1}),[\mathcal{M}^{2},\hat{\pi}(a^{0})]]...]
\end{equation}
For n$\geqslant 2$ we have 
\begin{equation}
\pi (J^{k+1}\frak{g)=}\left\{ \sigma (\omega ^{k});\omega ^{k}\in \hat{\Omega%
}^{k}\frak{g\cap }\ker (\pi )\right\} .
\end{equation}
The main difference between NCG and NAG is that in NAG the connection $\rho $
and the curvature $\theta $ are not, in general, elements of $\Omega ^{1}%
\frak{g}$ and $\Omega ^{2}\frak{g}$. To construct connection and curvature
we need the spaces $\mathbf{r}^{0}\frak{a}$ , $\mathbf{r}^{1}\frak{a}\subset
M_{F}(\Bbb{C})$ defined by the conditions:

\begin{eqnarray}
\frak{r}^{0}(\frak{a}) &=&\hat{\Gamma}\frak{r}^{0}(\frak{a})\hat{\Gamma}
\label{r0a} \\
\left[ \frak{r}^{0}(\frak{a}),\hat{\pi}(\frak{a})\right] &\subset &\hat{\pi}(%
\frak{a})  \nonumber \\
\left[ \frak{r}^{0}(\frak{a}),\hat{\pi}(\Omega ^{1}\frak{a})\right] &\subset
&\hat{\pi}(\Omega ^{1}\frak{a})  \nonumber \\
\left\{ \frak{r}^{0}(\frak{a}),\hat{\pi}(\frak{a})\right\} &\subset &\{\hat{%
\pi}(\frak{a}),\hat{\pi}(\frak{a})\}+\hat{\pi}(\Omega ^{2}\frak{a}) 
\nonumber \\
\left[ \frak{r}^{0}(\frak{a}),\hat{\pi}(\Omega ^{1}\frak{a})\right] &\subset
&\{\hat{\pi}(\frak{a}),\hat{\pi}(\Omega ^{1}\frak{a})\}+\hat{\pi}(\Omega ^{3}%
\frak{a})  \nonumber
\end{eqnarray}
and 
\begin{eqnarray}
\frak{r}^{1}(\frak{a}) &=&-\hat{\Gamma}\frak{r}^{1}(\frak{a})\hat{\Gamma}
\label{r1a} \\
\left[ \frak{r}^{1}(\frak{a}),\hat{\pi}(\frak{a})\right] &\subset &\hat{\pi}%
(\Omega ^{1}\frak{a})  \nonumber \\
\left\{ \frak{r}^{1}(\frak{a}),\hat{\pi}(\Omega ^{1}\frak{a})\right\}
&\subset &\hat{\pi}(\Omega ^{2}\frak{a})+\{\hat{\pi}(\frak{a}),\hat{\pi}(%
\frak{a})\}  \nonumber
\end{eqnarray}
where $\hat{\Gamma}$ is the grading operator.

The connection is given by: 
\begin{equation}
\rho =\stackunder{\alpha }{\sum }(c_{\alpha }^{1}\otimes m_{\alpha
}^{0}+c_{\alpha }^{0}\gamma ^{5}\otimes m_{\alpha }^{1})
\end{equation}
where 
\begin{equation}
c_{\alpha }^{1}\in \Lambda ^{1},c_{\alpha }^{0}\in \Lambda ^{0}
\end{equation}
$\Lambda ^{k}$ is the space of k-forms represented by gamma matrices, and 
\begin{eqnarray}
m_{\alpha }^{0} &\in &\frak{r}^{0}(\frak{a}) \\
m_{\alpha }^{1} &\in &\frak{r}^{1}(\frak{a})  \nonumber
\end{eqnarray}

To construct the curvature $\theta $ we need the spaces $\mathbf{j}^{0}\frak{%
a,}$ $\mathbf{j}^{1}\frak{a}$ and $\mathbf{j}^{2}\frak{a}$ defined as: 
\begin{equation}
\mathbf{j}^{0}\frak{a=}\mathbf{c}^{0}\frak{a}
\end{equation}
where 
\begin{eqnarray}
\mathbf{c}^{0}\frak{a}\text{ } &=&\mathbf{c}^{0}\frak{a}\text{ }\hat{\Gamma}%
\text{ }\mathbf{c}^{0}\frak{a}\text{ }  \label{c0a} \\
\mathbf{c}^{0}\frak{a}\text{ .}\hat{\pi}(a) &=&0  \nonumber \\
\mathbf{c}^{0}\frak{a}\text{ .}\hat{\pi}(\Omega ^{1}a) &=&0  \nonumber
\end{eqnarray}
and 
\begin{equation}
\mathbf{j}^{1}\frak{a=}\mathbf{c}^{1}\frak{a}
\end{equation}
where 
\begin{eqnarray}
\text{{}}\mathbf{c}^{1}\frak{a}\text{ } &=&-\mathbf{c}^{1}\frak{a}\text{ }%
\hat{\Gamma}\text{ }\mathbf{c}^{1}\frak{a}\text{ }  \label{c1a} \\
\mathbf{c}^{1}\frak{a}\text{ .}\hat{\pi}(a) &=&0  \nonumber \\
\mathbf{c}^{1}\frak{a}\text{ .}\hat{\pi}(\Omega ^{1}a) &=&0  \nonumber
\end{eqnarray}
and 
\begin{equation}
\mathbf{j}^{2}\frak{a=}\mathbf{c}^{2}\frak{a}+\hat{\pi}(\frak{J}^{2}\frak{a}%
)+\left\{ \hat{\pi}(\frak{a}),\hat{\pi}(\frak{a})\right\}
\end{equation}
with 
\[
\text{{}}\mathbf{c}^{2}\frak{a}\text{ }=\mathbf{c}^{2}\frak{a}\text{ }\hat{%
\Gamma}\text{ }\mathbf{c}^{2}\frak{a}\text{ } 
\]
\begin{eqnarray}
\left[ \mathbf{c}^{2}\frak{a}\text{ },\hat{\pi}(\frak{a})\right] &=&0
\label{c2a} \\
\left[ \mathbf{c}^{2}\frak{a}\text{ },\hat{\pi}(\Omega ^{1}\frak{a})\right]
&=&0  \nonumber
\end{eqnarray}

The curvature $\theta $ is given by: 
\begin{equation}
\theta =\mathbf{d}\rho +\rho ^{2}-i\{\gamma ^{5}\otimes \mathcal{M},\rho \}+%
\hat{\sigma}_{\frak{g}}(\rho )\gamma ^{5}+\Bbb{J}^{2}\frak{g}
\end{equation}
where 
\begin{equation}
\Bbb{J}^{2}\frak{g=}(\Lambda ^{2}\otimes {\footnotesize j}^{0}\frak{a)\oplus 
}(\Lambda ^{1}\gamma ^{5}\otimes {\footnotesize j}^{1}\frak{a)\oplus }%
(\Lambda ^{0}\otimes {\footnotesize j}^{2}\frak{a)}
\end{equation}

We have to choose the representative of the curvature $\theta $ , $%
\varepsilon (\theta ),$ orthogonal to $\Bbb{J}^{2}\frak{g,}$given by:

\begin{equation}
\varepsilon (\theta )=\mathbf{d}\rho +\rho ^{2}-i\{\gamma ^{5}\otimes 
\mathcal{M},\rho \}+\hat{\sigma}_{\frak{g}}(\rho )\gamma ^{5}+j
\end{equation}
where 
\begin{equation}
\int_{X}dxtr(\varepsilon (\theta )j^{\prime })=0~~\forall j^{\prime }\in 
\Bbb{J}^{2}\frak{g}
\end{equation}

The bosonic and the fermionic actions are given by: 
\begin{equation}
S_{B}=\frac{1}{N_{F}g_{0}^{2}}\int_{X}dxtr(\varepsilon (\theta )^{2})
\end{equation}
and 
\begin{equation}
S_{F}=\int_{X}\Psi ^{*}(D+i\rho )\Psi
\end{equation}
where $g_{0}$ is a coupling constant.

\section{The construction}

The discrete L-cycle ($\frak{a,}\mathcal{H},\mathcal{M})$ consists of the
Lie algebra given by: 
\begin{equation}
\frak{a=su(3)\oplus su(2)\oplus su(2)\oplus u(1)}
\end{equation}
\[
\ni \{a_{3},a_{2},a_{2}^{\prime },a_{1}\} 
\]

The Hilbert space $\mathcal{H}$ is $\Bbb{C}^{48}$ labelled by the elements 
\[
(\mathbf{u}_{L}\mathbf{,d}_{L}\mathbf{,u}_{R}\mathbf{,d}_{R},\nu
_{L},e_{L},\nu _{R},e_{R})^{T} 
\]
$\ $where $\mathbf{u}_{L}\mathbf{,d}_{L}\mathbf{,u}_{R}\mathbf{,d}_{R}\in 
\Bbb{C}^{3}\otimes \Bbb{C}^{3}$ and $\nu _{L},e_{L},\nu _{R},e_{R}\in \Bbb{C}%
^{3}.$

The Lie algebra $\frak{a}$ acts on $\mathcal{H}$ via the representation: 
\begin{equation}
\hat{\pi}((a_{1},a_{2},a_{2}^{\prime },a_{3}))=\left[ 
\begin{array}{cc}
\hat{\pi}_{q}((a_{1},a_{2},a_{2}^{\prime },a_{3})) & 0 \\ 
0 & \hat{\pi}_{l}((a_{1},a_{2},a_{2}^{\prime },a_{3}))
\end{array}
\right]
\end{equation}
where q and l label, respectively, the quarkionic and the leptonic sectors,
with the representation of the quark sector given by: 
\begin{equation}
\hat{\pi}_{q}((a_{1},a_{2},a_{2}^{\prime },a_{3}))=\left[ 
\begin{array}{cc}
\hat{\pi}_{q}^{1}((a_{1},a_{2},a_{2}^{\prime },a_{3})) & 0 \\ 
0 & \hat{\pi}_{q}^{2}((a_{1},a_{2},a_{2}^{\prime },a_{3}))
\end{array}
\right]
\end{equation}
where 
\begin{eqnarray}
\hat{\pi}_{q}^{1}((a_{1},a_{2},a_{2}^{\prime },a_{3})) &=&if_{0}diag(\alpha 
\mathbf{1}_{3}\otimes \mathbf{1}_{3},\beta \mathbf{1}_{3}\otimes \mathbf{1}%
_{3})+ \\
&&\left[ 
\begin{array}{cc}
(a_{3}+if_{3}^{1}\mathbf{1}_{3})\otimes \mathbf{1}_{3} & 
i(f_{1}^{1}-if_{2}^{1})\mathbf{1}_{3}\otimes \mathbf{1}_{3} \\ 
i(f_{1}^{1}+if_{2}^{1})\mathbf{1}_{3}\otimes \mathbf{1}_{3} & 
(a_{3}-if_{3}^{1}\mathbf{1}_{3})\otimes \mathbf{1}_{3}
\end{array}
\right]  \nonumber
\end{eqnarray}
and 
\begin{eqnarray}
\hat{\pi}_{q}^{2}((a_{1},a_{2},a_{2}^{\prime },a_{3})) &=&if_{0}diag(\gamma 
\mathbf{1}_{3}\otimes \mathbf{1}_{3},\delta \mathbf{1}_{3}\otimes \mathbf{1}%
_{3})+ \\
&&\left[ 
\begin{array}{cc}
(a_{3}+if_{3}^{2}\mathbf{1}_{3})\otimes \mathbf{1}_{3} & 
i(f_{1}^{2}-if_{2}^{2})\mathbf{1}_{3}\otimes \mathbf{1}_{3} \\ 
i(f_{1}^{2}+if_{2}^{2})\mathbf{1}_{3}\otimes \mathbf{1}_{3} & 
(a_{3}-if_{3}^{2}\mathbf{1}_{3})\otimes \mathbf{1}_{3}
\end{array}
\right]  \nonumber
\end{eqnarray}
and the representation of the leptonic sector: 
\begin{equation}
\hat{\pi}_{l}((a_{1},a_{2},a_{2}^{\prime },a_{3}))=\left[ 
\begin{array}{cc}
\hat{\pi}_{l}^{1}((a_{1},a_{2},a_{2}^{\prime },a_{3})) & 0 \\ 
0 & \hat{\pi}_{l}^{2}((a_{1},a_{2},a_{2}^{\prime },a_{3}))
\end{array}
\right]
\end{equation}
where

\begin{eqnarray}
\hat{\pi}_{l}^{1}((a_{1},a_{2},a_{2}^{\prime },a_{3}))
&=&if_{0}diag(\upsilon \mathbf{1}_{3},\varepsilon \mathbf{1}_{3})+  \nonumber
\\
&&\left[ 
\begin{array}{ll}
if_{3}^{1}\otimes \mathbf{1}_{3} & i(f_{1}^{1}-if_{2}^{1})\otimes \mathbf{1}%
_{3} \\ 
i(f_{1}^{1}+if_{2}^{1})\otimes \mathbf{1}_{3} & -if_{3}^{1}\otimes \mathbf{1}%
_{3}
\end{array}
\right]
\end{eqnarray}
and 
\begin{eqnarray}
\hat{\pi}_{l}^{2}((a_{1},a_{2},a_{2}^{\prime },a_{3})) &=&if_{0}diag(\zeta 
\mathbf{1}_{3},\eta \mathbf{1}_{3})+ \\
&&\left[ 
\begin{array}{ll}
if_{3}^{2}\otimes \mathbf{1}_{3} & i(f_{1}^{2}-if_{2}^{2})\otimes \mathbf{1}%
_{3} \\ 
i(f_{1}^{2}+if_{2}^{2})\otimes \mathbf{1}_{3} & -if_{3}^{2}\otimes \mathbf{1}%
_{3}
\end{array}
\right]  \nonumber
\end{eqnarray}
where we have used the fact that elements of $su(2)$ algebra have the
representation: 
\begin{equation}
a_{2}=\left[ 
\begin{array}{ll}
if_{3} & i(f_{1}-if_{2}) \\ 
i(f_{1}+if_{2}) & -if_{3}
\end{array}
\right] .
\end{equation}

The Dirac operator is given by: 
\begin{equation}
\mathcal{M=}\left[ 
\begin{array}{ll}
\mathcal{M}_{q} & 0 \\ 
0 & \mathcal{M}_{l}
\end{array}
\right]
\end{equation}
with 
\begin{equation}
\mathcal{M}_{q}=\left[ 
\begin{array}{cccc}
0 & 0 & \mathbf{1}_{3}\otimes M_{u} & \mathbf{0} \\ 
0 & 0 & 0 & \mathbf{1}_{3}\otimes M_{d} \\ 
\mathbf{1}_{3}\otimes M_{u}^{*} & 0 & 0 & 0 \\ 
0 & \mathbf{1}_{3}\otimes M_{d}^{*} & 0 & 0
\end{array}
\right]
\end{equation}
and 
\begin{equation}
\mathcal{M}_{l}=\left[ 
\begin{array}{cccc}
0 & 0 & M_{\nu } & \mathbf{0} \\ 
0 & 0 & 0 & M_{e} \\ 
M_{\nu }^{*} & 0 & 0 & 0 \\ 
0 & M_{e}^{*} & 0 & 0
\end{array}
\right]
\end{equation}
where $M_{u},M_{d},M_{\nu },M_{e}\in M_{3}(\Bbb{C})$ are the mass matrices
of the fermions.

Remark that we have chosen the coefficients associated with u(1) algebra
arbitrary, we will see when calculating the fermionic action that in fact
they corresponds to the hypercharges.

The grading operator is given by: 
\begin{equation}
\hat{\Gamma}=diag(-\mathbf{1}_{3}\otimes \mathbf{1}_{3},-\mathbf{1}%
_{3}\otimes \mathbf{1}_{3},\mathbf{1}_{3}\otimes \mathbf{1}_{3},\mathbf{1}%
_{3}\otimes \mathbf{1}_{3},-\mathbf{1}_{3},-\mathbf{1}_{3},\mathbf{1}_{3},%
\mathbf{1}_{3})
\end{equation}

The space $\hat{\pi}(\Omega ^{1}a)$ is generated by elements of the type

\begin{equation}
\tau ^{1}=\stackunder{\alpha ,z}{\sum }\left[ \hat{\pi}(a_{\alpha
}^{z}),...\left[ \hat{\pi}(a_{\alpha }^{1}),\left[ -i\mathcal{M},\hat{\pi}%
(a_{\alpha }^{0})\right] \right] ...\right]
\end{equation}
where $a_{\alpha }^{i}=(a_{1}^{i},a_{2}^{i},a_{2}^{i\prime },a_{3}^{i})$ .

Simple calculation gives the following decomposition: 
\begin{equation}
\tau ^{1}=\left[ 
\begin{array}{ll}
\tau _{q}^{1} & 0 \\ 
0 & \tau _{l}^{1}
\end{array}
\right]
\end{equation}
where for the quark sector we have 
\begin{equation}
\tau _{q}^{1}=\left[ 
\begin{array}{ll}
0 & \tau _{1q}^{1} \\ 
\tau _{2q}^{1} & 0
\end{array}
\right]
\end{equation}
with 
\begin{equation}
\tau _{1q}^{1}=i\left[ 
\begin{array}{ll}
\bar{b}_{2}1_{3}\otimes M_{d}+\bar{c}_{2}1_{3}\otimes M_{u} & 
b_{1}1_{3}\otimes M_{d}+c_{1}1_{3}\otimes M_{u} \\ 
-\bar{b}_{1}1_{3}\otimes M_{u}-\bar{c}_{1}1_{3}\otimes M_{d} & 
b_{2}1_{3}\otimes M_{u}+c_{2}1_{3}\otimes M_{d}
\end{array}
\right]
\end{equation}
and 
\begin{equation}
\tau _{2q}^{1}=i\left[ 
\begin{array}{ll}
b_{2}1_{3}\otimes M_{d}^{*}+c_{2}1_{3}\otimes M_{u}^{*} & -b_{1}1_{3}\otimes
M_{u}^{*}-c_{1}1_{3}\otimes M_{d}^{*} \\ 
\bar{b}_{1}1_{3}\otimes M_{d}^{*}+\bar{c}_{1}1_{3}\otimes M_{u}^{*} & \bar{b}%
_{2}1_{3}\otimes M_{u}^{*}+\bar{c}_{2}1_{3}\otimes M_{d}^{*}
\end{array}
\right] .
\end{equation}

For the leptonic sector we have 
\begin{equation}
\tau _{l}^{1}=\left[ 
\begin{array}{ll}
0 & \tau _{1l}^{1} \\ 
\tau _{2l}^{1} & 0
\end{array}
\right]
\end{equation}
where 
\begin{equation}
\tau _{1l}^{1}=i\left[ 
\begin{array}{ll}
\bar{b}_{2}\otimes M_{e}+\bar{c}_{2}\otimes M_{\nu } & b1\otimes
M_{e}+c_{1}\otimes M_{\nu } \\ 
-\bar{b}_{1}\otimes M_{\nu }-\bar{c}_{1}\otimes M_{e} & b_{2}\otimes M_{\nu
}+c_{2}\otimes M_{e}
\end{array}
\right]
\end{equation}
and 
\begin{equation}
\tau _{2l}^{1}=i\left[ 
\begin{array}{ll}
b_{2}\otimes M_{\nu }^{*}+c_{2}\otimes M_{e}^{*} & -b_{1}\otimes M_{\nu
}^{*}-c_{1}\otimes M_{e}^{*} \\ 
\bar{b}_{1}\otimes M_{e}^{*}+\bar{c}_{1}\otimes M_{\nu }^{*} & \bar{b}%
_{2}\otimes M_{e}^{*}+\bar{c}_{2}\otimes M_{\nu }^{*}
\end{array}
\right] .
\end{equation}

We can show that the form $\tau ^{1}$is antihermetian provided that: 
\begin{equation}
\alpha +\beta =\gamma +\delta
\end{equation}

and 
\begin{equation}
\upsilon +\varepsilon =\zeta +\eta .
\end{equation}

Similarly, straightforward calculations give: 
\begin{equation}
\hat{\pi}(\Omega ^{2}\frak{a})\ni \tau ^{2}=\{\tau ^{1},\tau ^{1}\}=\left[ 
\begin{array}{ll}
\tau _{q}^{2} & 0 \\ 
0 & \tau _{l}^{2}
\end{array}
\right]
\end{equation}
with 
\begin{equation}
\tau _{l}^{2}=\left[ 
\begin{array}{llll}
\tau _{11}^{l2} & \tau _{12}^{l2} & 0 & 0 \\ 
\tau _{21}^{l2} & \tau _{22}^{l2} & 0 & 0 \\ 
0 & 0 & \tau _{33}^{l2} & \tau _{34}^{l2} \\ 
0 & 0 & \tau _{43}^{l2} & \tau _{44}^{l2}
\end{array}
\right]
\end{equation}
where the diagonal elements are: 
\begin{eqnarray}
-\tau _{11}^{l2} &=&2(\left| b_{1}\right| ^{2}+\left| c_{2}\right|
^{2})\otimes M_{e}M_{e}^{*}+2(\left| c_{1}\right| ^{2}+\left| b_{2}\right|
^{2})\otimes M_{\nu }M_{\nu }^{*} \\
&&+2(b_{1}\bar{c}_{1}+\bar{c}_{2}b_{2})\otimes M_{e}M_{v}^{*}+2(\bar{b}%
_{1}c_{1}+c_{2}\bar{b}_{2})\otimes M_{\nu }M_{e}^{*}  \nonumber \\
-\tau _{22}^{l2} &=&2(\left| c_{1}\right| ^{2}+\left| b_{2}\right|
^{2})\otimes M_{e}M_{e}^{*}+2(\left| b_{1}\right| ^{2}+\left| c_{2}\right|
^{2})\otimes M_{\nu }M_{\nu }^{*}  \nonumber \\
&&+2(b_{1}\bar{c}_{1}+\bar{c}_{2}b_{2})\otimes M_{e}M_{v}^{*}+2(\bar{b}%
_{1}c_{1}+c_{2}\bar{b}_{2})\otimes M_{\nu }M_{e}^{*}  \nonumber \\
-\tau _{33}^{l2} &=&2(\left| c_{2}\right| ^{2}+\left| c_{1}\right|
^{2})\otimes M_{e}M_{e}^{*}+2(\left| b_{1}\right| ^{2}+\left| b_{2}\right|
^{2})\otimes M_{\nu }M_{\nu }^{*}  \nonumber \\
&&+2(b_{1}\bar{c}_{1}+\bar{c}_{2}b_{2})\otimes M_{e}M_{v}^{*}+2(\bar{b}%
_{1}c_{1}+c_{2}\bar{b}_{2})\otimes M_{\nu }M_{e}^{*}  \nonumber \\
-\tau _{44}^{l2} &=&2(\left| b_{1}\right| ^{2}+\left| b_{2}\right|
^{2})\otimes M_{e}M_{e}^{*}+2(\left| c_{1}\right| ^{2}+\left| c_{2}\right|
^{2})\otimes M_{\nu }M_{\nu }^{*}  \nonumber \\
&&+2(b_{1}\bar{c}_{1}+\bar{c}_{2}b_{2})\otimes M_{e}M_{v}^{*}+2(\bar{b}%
_{1}c_{1}+c_{2}\bar{b}_{2})\otimes M_{\nu }M_{e}^{*}  \nonumber
\end{eqnarray}
and the off-diagonal ones: 
\begin{eqnarray}
-\tau _{12}^{l2} &=&2(\bar{c}_{2}c_{1}-b_{1}\bar{b}_{2})\otimes M_{\nu e} \\
-\tau _{21}^{l2} &=&2(c_{2}\bar{c}_{1}-\bar{b}_{1}b_{2})\otimes M_{\nu e} 
\nonumber \\
-\tau _{34}^{l2} &=&2(c_{1}b_{2}-b_{1}c_{2})\otimes M_{\nu e}  \nonumber \\
-\tau _{43}^{l2} &=&2(\bar{c}_{1}\bar{b}_{2}-\bar{b}_{1}\bar{c}_{2})\otimes
M_{\nu e}  \nonumber
\end{eqnarray}
with 
\begin{eqnarray}
M_{\nu e} &=&M_{\nu }M_{\nu }^{*}-M_{e}M_{e}^{*} \\
M_{\{\nu e\}} &=&M_{\nu }M_{\nu }^{*}+M_{e}M_{e}^{*}  \nonumber
\end{eqnarray}
and similarly

\begin{equation}
\tau _{q}^{2}=\left[ 
\begin{array}{llll}
\tau _{11}^{q2} & \tau _{12}^{q2} & 0 & 0 \\ 
\tau _{21}^{q2} & \tau _{22}^{q2} & 0 & 0 \\ 
0 & 0 & \tau _{33}^{q2} & \tau _{34}^{q2} \\ 
0 & 0 & \tau _{43}^{q2} & \tau _{44}^{q2}
\end{array}
\right]
\end{equation}
with 
\begin{eqnarray}
-\tau _{11}^{q2} &=&2(\left| b_{1}\right| ^{2}+\left| c_{2}\right| ^{2})%
\mathbf{1}_{3}\otimes M_{d}M_{d}^{*}+2(\left| c_{1}\right| ^{2}+\left|
b_{2}\right| ^{2})\mathbf{1}_{3}\otimes M_{u}M_{u}^{*} \\
&&+2(b_{1}\bar{c}_{1}+\bar{c}_{2}b_{2})\mathbf{1}_{3}\otimes
M_{d}M_{u}^{*}+2(\bar{b}_{1}c_{1}+c_{2}\bar{b}_{2})\mathbf{1}_{3}\otimes
M_{u}M_{d}^{*}  \nonumber \\
-\tau _{22}^{l2} &=&2(\left| c_{1}\right| ^{2}+\left| b_{2}\right| ^{2})%
\mathbf{1}_{3}\otimes M_{d}M_{d}^{*}+2(\left| b_{1}\right| ^{2}+\left|
c_{2}\right| ^{2})\mathbf{1}_{3}\otimes M_{u}M_{u}^{*}  \nonumber \\
&&+2(b_{1}\bar{c}_{1}+\bar{c}_{2}b_{2})\mathbf{1}_{3}\otimes
M_{d}M_{u}^{*}+2(\bar{b}_{1}c_{1}+c_{2}\bar{b}_{2})\mathbf{1}_{3}\otimes
M_{u}M_{d}^{*}  \nonumber \\
-\tau _{33}^{l2} &=&2(\left| c_{2}\right| ^{2}+\left| c_{1}\right| ^{2})%
\mathbf{1}_{3}\otimes M_{d}M_{d}^{*}+2(\left| b_{1}\right| ^{2}+\left|
b_{2}\right| ^{2})\mathbf{1}_{3}\otimes M_{u}M_{u}^{*}  \nonumber \\
&&+2(b_{1}\bar{c}_{1}+\bar{c}_{2}b_{2})\mathbf{1}_{3}\otimes
M_{d}M_{u}^{*}+2(\bar{b}_{1}c_{1}+c_{2}\bar{b}_{2})\mathbf{1}_{3}\otimes
M_{u}M_{d}^{*}  \nonumber \\
-\tau _{44}^{l2} &=&2(\left| b_{1}\right| ^{2}+\left| b_{2}\right| ^{2})%
\mathbf{1}_{3}\otimes M_{d}M_{d}^{*}+2(\left| c_{1}\right| ^{2}+\left|
c_{2}\right| ^{2})\mathbf{1}_{3}\otimes M_{u}M_{u}^{*}  \nonumber \\
&&+2(b_{1}\bar{c}_{1}+\bar{c}_{2}b_{2})\mathbf{1}_{3}\otimes
M_{d}M_{u}^{*}+2(\bar{b}_{1}c_{1}+c_{2}\bar{b}_{2})\mathbf{1}_{3}\otimes
M_{u}M_{d}^{*}  \nonumber
\end{eqnarray}
and 
\begin{eqnarray}
-\tau _{12}^{q2} &=&2(\bar{c}_{2}c_{1}-b_{1}\bar{b}_{2})\mathbf{1}%
_{3}\otimes M_{ud} \\
-\tau _{21}^{q2} &=&2(c_{2}\bar{c}_{1}-\bar{b}_{1}b_{2})\mathbf{1}%
_{3}\otimes M_{ud}  \nonumber \\
-\tau _{34}^{q2} &=&2(c_{1}b_{2}-b_{1}c_{2})\mathbf{1}_{3}\otimes M_{ud} 
\nonumber \\
-\tau _{43}^{q2} &=&2(\bar{c}_{1}\bar{b}_{2}-\bar{b}_{1}\bar{c}_{2})\mathbf{1%
}_{3}\otimes M_{\nu ud}  \nonumber
\end{eqnarray}
where 
\begin{eqnarray}
M_{ud} &=&M_{u}M_{u}^{*}-M_{d}M_{d}^{*} \\
M_{\{ud\}} &=&M_{u}M_{u}^{*}+M_{d}M_{d}^{*}.  \nonumber
\end{eqnarray}

We can show that: 
\begin{equation}
\hat{\sigma}(\omega ^{1})=\left[ 
\begin{array}{ll}
\hat{\sigma}_{q}(\omega ^{1}) & 0 \\ 
0 & \hat{\sigma}_{l}(\omega ^{1})
\end{array}
\right]
\end{equation}
where

\begin{equation}
\hat{\sigma}_{q}(\omega ^{1})=\left[ 
\begin{array}{ll}
\hat{\sigma}_{1q} & 0 \\ 
0 & \hat{\sigma}_{2q}
\end{array}
\right]
\end{equation}
with 
\begin{equation}
\hat{\sigma}_{1q}=i\left[ 
\begin{array}{ll}
if_{3}^{1}1_{3}\otimes M_{ud} & i(f_{1}^{1}-if_{2}^{1})1_{3}\otimes M_{ud}
\\ 
i(f_{1}^{1}+if_{2}^{1})1_{3}\otimes M_{ud} & -if_{3}^{1}1_{3}\otimes M_{ud}
\end{array}
\right]
\end{equation}
and 
\begin{equation}
\hat{\sigma}_{2q}=i\left[ 
\begin{array}{ll}
if_{3}^{2}1_{3}\otimes M_{ud} & i(f_{1}^{2}-if_{2}^{2})1_{3}\otimes M_{ud}
\\ 
i(f_{1}^{2}+if_{2}^{2})1_{3}\otimes M_{ud} & -if_{3}^{2}1_{3}\otimes M_{ud}
\end{array}
\right] .
\end{equation}

For the leptonic sector we have

\begin{equation}
\hat{\sigma}_{l}(\omega ^{1})=\left[ 
\begin{array}{ll}
\hat{\sigma}_{1l} & 0 \\ 
0 & \hat{\sigma}_{2l}
\end{array}
\right]
\end{equation}
with 
\begin{equation}
\hat{\sigma}_{1l}=i\left[ 
\begin{array}{ll}
if_{3}^{1}\otimes M_{\nu e} & i(f_{1}^{1}-if_{2}^{1})\otimes M_{\nu e} \\ 
i(f_{1}^{1}+if_{2}^{1})\otimes M_{\nu e} & -if_{3}^{1}\otimes M_{\nu e}
\end{array}
\right]
\end{equation}
and 
\begin{equation}
\hat{\sigma}_{2l}=i\left[ 
\begin{array}{ll}
if_{3}^{2}\otimes M_{\nu e} & i(f_{1}^{2}-if_{2}^{2})\otimes M_{\nu e} \\ 
i(f_{1}^{2}+if_{2}^{2})\otimes M_{\nu e} & -if_{3}^{2}\otimes M_{\nu e}
\end{array}
\right] .
\end{equation}

After trivial calculation we can show that

\begin{eqnarray}
\tau ^{2} &=&diag(K^{\prime }\otimes 1_{3},K^{\prime }\otimes
1_{3},K^{\prime }\otimes 1_{3},K^{\prime }\otimes 1_{3},  \nonumber \\
&&K,K,K,K)\func{mod}(\hat{\sigma}(\Omega ^{1}a))
\end{eqnarray}
with 
\begin{equation}
K=\hat{\alpha}\text{ }M_{\{\nu e\}}+\hat{\beta}\text{ }M_{e}M_{v}^{*}+\hat{%
\gamma}\text{ }M_{\nu }M_{e}^{*}
\end{equation}
and 
\begin{equation}
K^{\prime }=\hat{\alpha}\text{ }M_{\{ud\}}+\hat{\beta}\text{ }M_{d}M_{u}^{*}+%
\hat{\gamma}\text{ }M_{u}M_{d}^{*}
\end{equation}
where 
\begin{eqnarray}
\hat{\alpha} &=&\left| b_{1}\right| ^{2}+\left| b_{2}\right| ^{2}+\left|
c_{1}\right| ^{2}+\left| c_{2}\right| ^{2} \\
\hat{\beta} &=&b_{1}\bar{c}_{1}+b_{2}\bar{c}_{2}  \nonumber \\
\hat{\gamma} &=&\stackrel{\bar{\wedge}}{\beta }=\bar{b}_{1}c_{1}+\bar{b}%
_{2}c_{2}.  \nonumber
\end{eqnarray}

Using the conditions given by eqs (\ref{r0a},\ref{r1a}), we can show that:

\begin{eqnarray}
\frak{r}^{0}(\frak{a}) &=&\hat{\pi}(\frak{a}) \\
\frak{r}^{1}(\frak{a}) &=&\hat{\pi}(\Omega ^{1}\frak{a})  \nonumber
\end{eqnarray}

Let us adopt the following parametrization

\begin{equation}
\stackunder{\alpha }{\sum }c_{\alpha }^{1}\otimes a_{1\alpha }=\stackunder{%
\alpha }{\sum }c_{\alpha }^{1}\otimes if_{0\alpha }=i\mathbf{\tilde{A}}\in
\Lambda ^{1}\otimes u(1)=i\frac{g^{\prime }}{2}\gamma ^{\mu }C_{\mu }
\end{equation}

\begin{equation}
\stackunder{\alpha }{\sum }c_{\alpha }^{1}\otimes a_{3\alpha }=\mathbf{\hat{A%
}}\in \Lambda ^{1}\otimes su(3)
\end{equation}

\begin{eqnarray}
\stackunder{\alpha }{\sum }c_{\alpha }^{1}\otimes a_{2\alpha } &=&A\in
\Lambda ^{1}\otimes su(2) \\
&=&\left[ 
\begin{array}{ll}
\stackunder{\alpha }{\sum }c_{\alpha }^{1}\otimes if_{3\alpha } & 
\stackunder{\alpha }{\sum }c_{\alpha }^{1}\otimes i(f_{1\alpha }-if_{2\alpha
}) \\ 
\stackunder{\alpha }{\sum }c_{\alpha }^{1}\otimes i(f_{1\alpha }+if_{2\alpha
}) & \stackunder{\alpha }{\sum }c_{\alpha }^{1}\otimes (-if_{3\alpha })
\end{array}
\right]  \nonumber \\
&=&\left[ 
\begin{array}{ll}
iA_{3} & i(A_{1}-iA_{2}) \\ 
i(A_{1}+iA_{2}) & -iA_{3}
\end{array}
\right]  \nonumber
\end{eqnarray}

and 
\begin{eqnarray}
\Phi _{i} &=&-\stackunder{\alpha }{\sum }c_{\alpha }^{0}\otimes b_{i\alpha
}\in \Lambda ^{0}\otimes \Bbb{C} \\
\Xi _{i} &=&-\stackunder{\alpha }{\sum }c_{\alpha }^{0}\otimes c_{i\alpha
}\in \Lambda ^{0}\otimes \Bbb{C}  \nonumber
\end{eqnarray}
The connection $\rho $ is given by 
\begin{equation}
\rho =\left[ 
\begin{array}{ll}
\rho _{q} & 0 \\ 
0 & \rho _{l}
\end{array}
\right]  \label{rho}
\end{equation}
where

\begin{equation}
\rho _{l}=\left[ 
\begin{array}{ll}
\rho _{l}^{A_{1}} & \rho _{l}^{H_{1}} \\ 
\rho _{l}^{H2} & \rho _{l}^{A_{2}}
\end{array}
\right]
\end{equation}
with 
\begin{eqnarray}
\rho _{l}^{A_{1}} &=&\left[ 
\begin{array}{ll}
i(\upsilon \tilde{A}+A_{3})\otimes 1_{3} & i(A_{1}-iA_{2})\otimes 1_{3} \\ 
i(A_{1}+iA_{2})\otimes 1_{3} & i(\varepsilon \tilde{A}-A_{3})\otimes 1_{3}
\end{array}
\right] \\
\rho _{l}^{A_{2}} &=&\left[ 
\begin{array}{ll}
i(\zeta \tilde{A}+A_{3}^{\prime })\otimes 1_{3} & i(A_{1}^{\prime
}-iA_{2}^{\prime })\otimes 1_{3} \\ 
i(A_{1}^{\prime }+iA_{2}^{\prime })\otimes 1_{3} & i(\eta \tilde{A}%
-A_{3}^{\prime })\otimes 1_{3}
\end{array}
\right]  \nonumber \\
\rho _{l}^{H_{1}} &=&\left[ 
\begin{array}{ll}
-i\gamma ^{5}\bar{\Phi}_{2}\otimes M_{e}-i\gamma ^{5}\bar{\Xi}_{2}\otimes
M_{\nu } & -i\gamma ^{5}\Phi _{1}\otimes M_{e}-i\gamma ^{5}\Xi _{1}\otimes
M_{\nu } \\ 
i\gamma ^{5}\bar{\Phi}_{1}1_{3}\otimes M_{\nu }+i\gamma ^{5}\bar{\Xi}%
_{1}\otimes M_{e} & -i\gamma ^{5}\Phi _{2}\otimes M_{\nu }-i\gamma ^{5}\Xi
_{2}\otimes M_{e}
\end{array}
\right]  \nonumber \\
\rho _{l}^{H2} &=&\left[ 
\begin{array}{ll}
-i\gamma ^{5}\Phi _{2}\otimes M_{e}^{*}-i\gamma ^{5}\Xi _{2}\otimes M_{\nu
}^{*} & i\gamma ^{5}\Phi _{1}\otimes M_{\nu }^{*}+i\gamma ^{5}\Xi
_{1}\otimes M_{e}^{*} \\ 
-i\gamma ^{5}\bar{\Phi}_{1}\otimes M_{e}^{*}-i\gamma ^{5}\bar{\Xi}%
_{1}\otimes M_{\nu }^{*} & -i\gamma ^{5}\bar{\Phi}_{2}\otimes M_{\nu
}^{*}-i\gamma ^{5}\bar{\Xi}_{2}\otimes M_{e}^{*}
\end{array}
\right]  \nonumber
\end{eqnarray}
and

\begin{equation}
\rho _{q}=\left[ 
\begin{array}{ll}
\rho _{q}^{A_{1}} & \rho _{q}^{H_{1}} \\ 
\rho _{q}^{H2} & \rho _{q}^{A_{2}}
\end{array}
\right]
\end{equation}
with 
\[
\rho _{q}^{A_{1}}=\left[ 
\begin{array}{ll}
(\hat{A}+i(\alpha \tilde{A}+A_{3})1_{3})\otimes 1_{3} & i(A_{1}-iA_{2})1_{3}%
\otimes 1_{3} \\ 
i(A_{1}+iA_{2})1_{3}\otimes 1_{3} & (\hat{A}+i(\beta \tilde{A}%
-A_{3})1_{3})\otimes 1_{3}
\end{array}
\right] 
\]

\begin{equation}
\rho _{q}^{A_{2}}=\left[ 
\begin{array}{ll}
(\hat{A}+i(\gamma \tilde{A}+A_{3}^{\prime })1_{3})\otimes 1_{3} & 
i(A_{1}^{\prime }-iA_{2}^{\prime })1_{3}\otimes 1_{3} \\ 
i(A_{1}^{\prime }+iA_{2}^{\prime })1_{3}\otimes 1_{3} & (\hat{A}+i(\delta 
\tilde{A}-A_{3}^{\prime })1_{3})\otimes 1_{3}
\end{array}
\right]
\end{equation}
\[
\rho _{q}^{H_{1}}=\left[ 
\begin{array}{ll}
-i\gamma ^{5}\bar{\Phi}_{2}1_{3}\otimes M_{d}-i\gamma ^{5}\bar{\Xi}%
_{2}1_{3}\otimes M_{u} & -i\gamma ^{5}\Phi _{1}1_{3}\otimes M_{d}-i\gamma
^{5}\Xi _{1}1_{3}\otimes M_{u} \\ 
i\gamma ^{5}\bar{\Phi}_{1}1_{3}\otimes M_{u}+i\gamma ^{5}\bar{\Xi}%
_{1}1_{3}\otimes M_{d} & -i\gamma ^{5}\Phi _{2}1_{3}\otimes M_{u}-i\gamma
^{5}\Xi _{2}1_{3}\otimes M_{d}
\end{array}
\right] 
\]
\[
\rho _{q}^{H_{2}}=\left[ 
\begin{array}{ll}
-i\gamma ^{5}\Phi _{2}1_{3}\otimes M_{d}^{*}-i\gamma ^{5}\Xi
_{2}1_{3}\otimes M_{e}^{*} & i\gamma ^{5}\Phi _{1}1_{3}\otimes
M_{u}^{*}+i\gamma ^{5}\Xi _{1}1_{3}\otimes M_{d}^{*} \\ 
-i\gamma ^{5}\bar{\Phi}_{1}1_{3}\otimes M_{d}^{*}-i\gamma ^{5}\bar{\Xi}%
_{1}1_{3}\otimes M_{e}^{*} & -i\gamma ^{5}\bar{\Phi}_{2}1_{3}\otimes
M_{u}^{*}-i\gamma ^{5}\bar{\Xi}_{2}1_{3}\otimes M_{d}^{*}
\end{array}
\right] 
\]

The interaction fermions-bosons is given by 
\begin{equation}
\frak{L}_{F-B}=i\Psi ^{*}\rho \Psi =\frak{L}^{lB}+\frak{L}^{qB}+\frak{L}%
^{lH}+\frak{L}^{qH},
\end{equation}
where $\frak{L}^{lB},\frak{L}^{qB};\frak{L}^{lH}$ , $\frak{L}^{qH}$ are
,respectively, the Lagrangians of interaction leptons-bosons, quarks-bosons,
leptons-Higgs and quark-Higgs, and 
\begin{equation}
\Psi =(\mathbf{u}_{L},\mathbf{d}_{L},\mathbf{u}_{R},\mathbf{d}_{R},\nu
_{L},e_{L},\nu _{R},e_{R})^{T}.
\end{equation}

Using the parametrization

\begin{equation}
\mathbf{\tilde{A}}=\mathbf{i}\frac{g^{\prime }}{2}\gamma ^{\mu }C_{\mu }
\end{equation}

\begin{equation}
\mathbf{A}=i\frac{g_{L}}{2}\stackunder{a=1}{\stackrel{3}{\sum }}W_{\mu
}^{a}\gamma ^{\mu }\otimes \sigma ^{a}
\end{equation}

\begin{equation}
\mathbf{A}^{\prime }=i\frac{g_{R}}{2}\stackunder{a=1}{\stackrel{3}{\sum }}%
W_{\mu }^{\prime a}\gamma ^{\mu }\otimes \sigma ^{a}
\end{equation}

and 
\begin{equation}
\mathbf{\hat{A}=}i\frac{g_{s}}{2}\stackunder{a=1}{\stackrel{3}{\sum }}G_{\mu
}^{a}\gamma ^{\mu }\otimes \lambda ^{a}
\end{equation}

where $g^{\prime },g_{L}$, $g_{R}$, $g_{s}$ are respectively the coupling
constants of $u(1)_{L-B},$ $su(2)_{L}$, $su(2)_{R}$ and $su(3)$ algebras.

Let \cite{cuypers} 
\begin{eqnarray}
B_{\mu } &=&\cos (\theta _{W})A_{\mu }-\sin (\theta _{W})Z_{\mu } \\
W_{\mu }^{3} &=&\sin (\theta _{W})A_{\mu }+\cos (\theta _{W})Z_{\mu } 
\nonumber
\end{eqnarray}
and

\begin{eqnarray}
C_{\mu } &=&\cos (\theta _{S})B_{\mu }-\sin (\theta _{S})Z_{\mu }^{\prime }
\\
W_{\mu }^{\prime 3} &=&\sin (\theta _{S})B_{\mu }+\cos (\theta _{S})Z_{\mu
}^{\prime },  \nonumber
\end{eqnarray}
where the field $B_{\mu }$ is associated to the group $U(1)_{Y},$ and $%
\theta _{W}$ and $\theta _{S}$ are the mixing angles. Once performing a Wick
rotation and by identification with the fermionic action provided by the
classical model we obtain \cite{Benslama}:

\begin{equation}
\varepsilon =\eta =\upsilon =\zeta =-1
\end{equation}
which are the leptons hypercharges in the left-right model \cite{Okumura}.

The same treatment, applied to the quarkionic sector, gives

\begin{equation}
\alpha =\beta =\gamma =\delta =\frac{1}{3}
\end{equation}
which are the quark hypercharges.

Using the conditions given by eqs. (\ref{c0a}, \ref{c1a} and \ref{c2a}) we
can show that 
\begin{equation}
\mathbf{j}^{0}\frak{a}=0
\end{equation}

\begin{equation}
\mathbf{j}^{1}\frak{a}=0
\end{equation}
and 
\begin{equation}
\mathbf{j}^{2}\frak{a=}diag(\Bbb{R}\mathbf{I}_{36},\Bbb{R}\mathbf{I}_{12})+%
\hat{\pi}(\frak{J}^{2}\frak{a})+\left\{ \hat{\pi}(\frak{a}),\hat{\pi}(\frak{a%
})\right\} .
\end{equation}

Let the anticommutator 
\begin{equation}
\hat{\psi}(\frak{a})=\left\{ \hat{\pi}(\frak{a}),\hat{\pi}(\frak{a})\right\}
=diag(\left\{ \hat{\pi}_{q}(\frak{a}),\hat{\pi}_{q}(\frak{a})\right\}
,\left\{ \hat{\pi}_{l}(\frak{a}),\hat{\pi}_{l}(\frak{a})\right\} )
\end{equation}
where

\begin{equation}
\left\{ \hat{\pi}_{q}(a),\hat{\pi}_{q}(a)\right\} \ni A_{q}+\Delta _{q}
\end{equation}
with 
\begin{equation}
A_{q}=\left[ 
\begin{array}{ll}
A_{q}^{1} & 0 \\ 
0 & A_{q}^{2}
\end{array}
\right] {\small \otimes I}_{3}
\end{equation}
where 
\begin{eqnarray}
A_{q}^{1} &=&i\left[ 
\begin{array}{ll}
\lambda _{1}^{0}a_{3}+i\theta _{1}^{0}I_{3} & (\varpi ^{1}-i\varpi
^{2})a_{3}+i(\hat{\lambda}^{1}-i\hat{\lambda}^{2})I_{3} \\ 
(\varpi ^{1}+i\varpi ^{2})a_{3}+i(\hat{\lambda}^{1}+i\hat{\lambda}^{2})I_{3}
& \lambda _{2}^{0}a_{3}-i\theta _{1}^{0}I_{3}
\end{array}
\right]  \nonumber \\
&&
\end{eqnarray}
\begin{eqnarray}
A_{q}^{2} &=&i\left[ 
\begin{array}{ll}
\lambda _{3}^{0}a_{3}+i\theta _{2}^{0}I_{3} & (\varpi ^{3}-i\varpi
^{4})a_{3}+i(\hat{\lambda}^{3}-i\hat{\lambda}^{4})I_{3} \\ 
(\varpi ^{3}+i\varpi ^{4})a_{3}+i(\hat{\lambda}^{3}+i\hat{\lambda}^{4})I_{3}
& \lambda _{4}^{0}a_{3}-i\theta _{2}^{0}I_{3}
\end{array}
\right]  \nonumber \\
&&
\end{eqnarray}
and

\begin{equation}
\Delta _{q}=diag(\lambda _{1}\mathbf{I}_{3},\lambda _{1}\mathbf{I}%
_{3},\lambda _{2}\mathbf{I}_{3},\lambda _{2}\mathbf{I}_{3})\otimes \mathbf{I}%
_{3}
\end{equation}
with the condition 
\begin{equation}
\lambda _{1}^{0}+\lambda _{2}^{0}=\lambda _{3}^{0}+\lambda _{4}^{0}
\end{equation}
and

\begin{equation}
\left\{ \hat{\pi}_{l}(a),\hat{\pi}_{l}(a)\right\} \ni A_{l}+\Delta _{l}
\end{equation}
with

\begin{equation}
A_{l}=i\left[ 
\begin{array}{llll}
i\kappa _{1}^{0}I_{3} & i(\check{\lambda}^{1}-i\check{\lambda}^{2})I_{3} & 0
& 0 \\ 
i(\check{\lambda}^{1}+i\check{\lambda}^{2})I_{3} & -i\kappa _{1}^{0}I_{3} & 0
& 0 \\ 
0 & 0 & i\kappa _{2}^{0}I_{3} & i(\check{\lambda}^{3}-i\check{\lambda}%
^{4})I_{3} \\ 
0 & 0 & i(\check{\lambda}^{3}+i\check{\lambda}^{4})I_{3} & -i\kappa
_{2}^{0}I_{3}
\end{array}
\right]
\end{equation}
and 
\begin{equation}
\Delta _{l}=diag(\chi \mathbf{I}_{3},\chi \mathbf{I}_{3},\chi \mathbf{I}%
_{3},\chi \mathbf{I}_{3})
\end{equation}
with $\lambda _{i},\lambda _{i}^{0},\check{\lambda}^{i},\hat{\lambda}%
^{i},\theta _{j}^{0},\kappa _{j}^{0},\varpi ^{i},\chi $ ($i=1..4,$ $j=1..2)$ 
$\in \Bbb{R}.$

Trivial calculations give \cite{Benslama}

\begin{equation}
\mathbf{j}^{2}a=\hat{\pi}(\mathcal{J}^{2}\frak{a})\oplus
diag(A_{q},A_{l})\oplus diag(K_{q},K_{l})
\end{equation}
where 
\begin{eqnarray}
K_{q} &=&diag(\varphi _{1}\mathbf{I}_{3},\varphi _{1}\mathbf{I}_{3},\varphi
_{2}\mathbf{I}_{3},\varphi _{2}\mathbf{I}_{3})\otimes \mathbf{I}_{3} \\
K_{l} &=&diag(\varphi _{3}\mathbf{I}_{3},\varphi _{3}\mathbf{I}_{3},\varphi
_{3}\mathbf{I}_{3},\varphi _{3}\mathbf{I}_{3})  \nonumber
\end{eqnarray}
with $\varphi _{1},\varphi _{2}$ et $\varphi _{3}\in \Bbb{R}.$

The representative $\varepsilon (\left\{ \tau ^{1},\tau ^{1}\right\} )$ of $%
\left\{ \tau ^{1},\tau ^{1}\right\} $ + $\mathbf{j}^{2}a$ orthogonal to $%
\mathbf{j}^{2}a$ is given by:

\begin{eqnarray}
\varepsilon {\tiny (}\left\{ \tau ^{1},\tau ^{1}\right\} {\tiny )} &=&diag(%
\tilde{K}^{\prime }\otimes 1_{3},\tilde{K}^{\prime }\otimes 1_{3},\tilde{K}%
^{\prime }\otimes 1_{3},\tilde{K}^{\prime }\otimes 1_{3},  \nonumber \\
&&\tilde{K},\tilde{K},\tilde{K},\tilde{K})\func{mod}(\hat{\sigma}(\Omega
^{1}a))
\end{eqnarray}
with 
\begin{eqnarray}
\tilde{K} &=&\hat{\alpha}\text{ }\tilde{M}_{\{\nu e\}}+\hat{\beta}\text{ }%
\stackrel{\backsim }{M_{e}M_{v}^{*}}+\hat{\gamma}\text{ }\stackrel{\backsim 
}{M_{\nu }M_{e}^{*}} \\
\tilde{K}^{\prime } &=&\hat{\alpha}\text{ }\tilde{M}_{\{ud\}}+\hat{\beta}%
\text{ }\stackrel{\backsim }{M_{d}M_{u}^{*}}+\hat{\gamma}\text{ }\stackrel{%
\backsim }{M_{u}M_{d}^{*}}  \nonumber
\end{eqnarray}
where 
\[
\tilde{M}_{\{\nu e\}}=M_{\nu }M_{\nu }^{*}+M_{e}M_{e}^{*}-\frac{1}{3}%
tr(M_{\nu }M_{\nu }^{*}+M_{e}M_{e}^{*})\mathbf{1}_{3}, 
\]

\[
\stackrel{\backsim }{M_{e}M_{v}^{*}}=M_{e}M_{v}^{*}-\frac{1}{3}%
tr(M_{e}M_{v}^{*})\mathbf{1}_{3}, 
\]

\begin{equation}
\stackrel{\backsim }{M_{\nu }M_{e}^{*}}=M_{\nu }M_{e}^{*}-\frac{1}{3}%
tr(M_{\nu }M_{e}^{*})\mathbf{1}_{3},
\end{equation}

\[
\text{ }\tilde{M}_{\{ud\}}=\text{ }=M_{u}M_{u}^{*}+M_{d}M_{d}^{*}-\frac{1}{3}%
tr(M_{u}M_{u}^{*}+M_{d}M_{d}^{*})\mathbf{1}_{3,} 
\]

\[
\stackrel{\backsim }{M_{d}M_{u}^{*}}=M_{d}M_{u}^{*}-\frac{1}{3}%
tr(M_{d}M_{u}^{*})\mathbf{1}_{3,} 
\]
and 
\[
\stackrel{\backsim }{M_{u}M_{d}^{*}}=M_{u}M_{d}^{*}-\frac{1}{3}%
tr(M_{u}M_{d}^{*})\mathbf{1}_{3}. 
\]

Remark also that

\begin{eqnarray}
\mathcal{M}^{2} &=&\frac{1}{2}diag(1_{3}\otimes \tilde{M}_{\{ud\}},1_{3}%
\otimes \tilde{M}_{\{ud\}},1_{3}\otimes \tilde{M}_{\{ud\}},1_{3}\otimes 
\tilde{M}_{\{ud\}},  \nonumber \\
&&\tilde{M}_{\{\nu e\}},\tilde{M}_{\{\nu e\}},\tilde{M}_{\{\nu e\}},\tilde{M}%
_{\{\nu e\}})\func{mod}(\hat{\sigma}(\Omega ^{1}a))
\end{eqnarray}

The bosonic action is given by 
\begin{equation}
S_{B}=\int_{X}(\frak{L}_{0}+\frak{L}_{1}+\frak{L}_{2})dx
\end{equation}
where 
\[
\frak{L}_{0}=\frac{1}{96g_{0}^{2}}\{tr(12\tilde{M}_{\{ud\}}^{2}+4\tilde{M}%
_{\{\nu e\}}^{2})\times (\left| \Phi _{1}\right| ^{2}{\tiny +}\left| \Phi
_{2}\right| ^{2}{\tiny +}\left| \Xi _{1}\right| ^{2}{\tiny +}\left| \Xi
_{2}+1\right| ^{2}-1)^{2} 
\]

\[
+tr(\stackrel{\backsim }{12(M_{d}M_{u}^{*})^{2}}+\stackrel{\backsim }{%
4(M_{e}M_{\nu }^{*})^{2}})\times (\Phi _{1}\stackrel{\_}{\Xi }_{1}+\Phi _{2}(%
\stackrel{\_}{\Xi }_{2}+1)) 
\]
\begin{equation}
+tr(\stackrel{\backsim }{12(M_{u}M_{d}^{*})^{2}}+\stackrel{\backsim }{%
4(M_{\nu }M_{e}^{*})^{2}})\times (\stackrel{\_}{\Phi }_{1}\Xi _{1}+\stackrel{%
\_}{\Phi }_{2}(\Xi _{2}+1))\}  \label{h}
\end{equation}

is the Higgs Lagrangian,

and

\begin{eqnarray}
\frak{L}_{2} &=&\frac{1}{4g_{0}^{2}}tr_{c}(\mathbf{d}\text{ }\mathbf{\hat{A}+%
}\frac{1}{2}\left\{ \mathbf{\hat{A},\hat{A}}\right\} )^{2}+\frac{1}{%
4g_{0}^{2}}tr_{c}((\mathbf{d}\text{ }\mathbf{A+}\frac{1}{2}\left\{ \mathbf{%
A,A}\right\} )^{2})  \label{free} \\
&&+\frac{1}{4g_{0}^{2}}tr_{c}(\mathbf{d}\text{ }\mathbf{A}^{\prime }\mathbf{+%
}\frac{1}{2}\left\{ \mathbf{A}^{\prime }\mathbf{,A}^{\prime }\right\} )^{2}+%
\frac{1}{3g_{0}^{2}}tr_{c}((\mathbf{d}\text{ }\mathbf{\tilde{A}}\text{ }%
)^{2}))  \nonumber
\end{eqnarray}
is the Yang-Mills Lagrangian.

Since $\frak{L}_{1},$ the Lagrangian of interaction Higgs-bosons, is quite
long it is not displayed here in its totality \cite{Benslama}. The parts of $%
\frak{L}_{1}$ which give rise to the masses of $W_{R}$ and $W_{L}$ bosons
are respectively: 
\begin{equation}
\mathcal{L}_{1}^{1}=\frac{1}{8g_{0}^{2}}\mathbf{m}^{2}tr_{c}\left|
i(A_{1}-iA_{2})(\Xi _{2}+1)\right| ^{2}
\end{equation}
and 
\begin{equation}
\mathcal{L}_{1}^{2}=\frac{1}{8g_{0}^{2}}\mathbf{m}^{2}tr_{c}\left|
i(A_{1}^{\prime }-iA_{2}^{\prime })(\Xi _{2}+1)\right| ^{2},
\end{equation}
where 
\begin{equation}
\mathbf{m}^{2}=tr(\frac{1}{3}M_{e}M_{e}^{*}+\frac{1}{3}M_{\nu }M_{\nu
}^{*}+M_{u}M_{u}^{*}+M_{d}M_{d}^{*}).
\end{equation}
Let us make the following rescaling 
\begin{equation}
\Xi _{2}=\frac{\bar{g}\phi _{2}}{\mathbf{m}},
\end{equation}
where $\bar{g}$ is a coupling constant which is specified later.

Since 
\begin{eqnarray}
i(A_{1}-iA_{2}) &=&i\frac{g_{L}}{\sqrt{2}}W_{\mu L}^{\dagger }\gamma ^{\mu }
\\
i(A_{1}^{\prime }-iA_{2}^{\prime }) &=&i\frac{g_{L}}{\sqrt{2}}W_{\mu
R}^{\dagger }\gamma ^{\mu }  \nonumber
\end{eqnarray}
and 
\begin{equation}
tr_{c}(\gamma ^{\mu }\gamma ^{\nu })=4\delta ^{\mu \nu }
\end{equation}
then 
\begin{equation}
\mathcal{L}_{1}^{1}=\frac{1}{4}(\frac{\bar{g}g_{L}}{g_{0}})^{2}W_{\mu
L}^{\dagger }W_{L}^{\mu }\phi _{2}^{2}+\frac{1}{2}(\frac{g_{L}}{g_{0}})^{2}%
\mathbf{m}\bar{g}W_{\mu L}^{\dagger }W_{L}^{\mu }\phi _{2}+\frac{1}{4}(\frac{%
g_{L}}{g_{0}})^{2}\mathbf{m}^{2}W_{\mu L}^{\dagger }W_{L}^{\mu }
\end{equation}

The last term gives:

\begin{equation}
M_{W_{L}}=\frac{1}{2}(\frac{g_{L}}{g_{0}})\mathbf{m}
\end{equation}

We can show also that: 
\begin{equation}
M_{W_{R}}=\frac{1}{2}(\frac{g_{R}}{g_{0}})\mathbf{m}
\end{equation}

Hence 
\begin{equation}
\frac{M_{W_{L}}}{M_{W_{R}}}=\frac{g_{L}}{g_{R}}
\end{equation}

In the case of left-right symmetric model 
\begin{equation}
g_{L}=g_{R,}
\end{equation}
we have

\begin{equation}
M_{W_{L}}=M_{W_{R}.}
\end{equation}
and the left-handed and the right-handed gauge bosons acquire the same
masses, leading to the conclusion that parity violation does not occur for
such models (\cite{Benslama1},\cite{Benslama2}). The same conclusion has
been obtained by \cite{Iochum} when discussing the $U(2)_{L}\times U(2)_{R}$
model in the framework of noncommutative geometry \`{a} la Connes.

For left-right symmetric model the nonassociative geometry approach does not
accommodate parity violation to be contrasted to Coquereaux approach \cite
{Okumura}.

Remark that if we neglect the mass of fermions in respect of the mass of the
top quark $m_{t}$ we obtain

\begin{equation}
M_{W_{L}}=\frac{1}{2}(\frac{g_{L}}{g_{0}})m_{t}
\end{equation}

and 
\begin{equation}
M_{W_{R}}=\frac{1}{2}(\frac{g_{R}}{g_{0}})m_{t}.
\end{equation}

The masses $Z$ et $Z^{\prime }$ bosons come from 
\begin{equation}
\mathcal{L}_{1}^{3}=\frac{1}{8g_{0}^{2}}\mathbf{m}^{2}tr_{c}\left|
i(A_{0}-iA_{3})(\Xi _{2}+1)\right| ^{2}
\end{equation}
and 
\begin{equation}
\mathcal{L}_{1}^{4}=\frac{1}{8g_{0}^{2}}\mathbf{m}^{2}tr_{c}\left|
i(A_{0}-iA_{3}^{\prime })(\Xi _{2}+1)\right| ^{2}.
\end{equation}

The calculations give 
\begin{equation}
M_{Z}=\frac{\mathbf{m}}{2}(\frac{g_{L}}{g_{0}})\frac{1}{\cos (\theta _{W})}.
\end{equation}

Since 
\begin{equation}
M_{W_{L}}=\frac{1}{2}(\frac{g_{L}}{g_{0}})\mathbf{m}
\end{equation}
then 
\begin{equation}
M_{Z}=\frac{M_{W_{L}}}{\cos (\theta _{W})},
\end{equation}
which is the same relation as in the standard model.

Y. Okumura \cite{Okumura} has found the same relation when applying
noncommutative geometry \`{a} la Coquereaux to left-right gauge model.

Similarly, we obtain 
\begin{equation}
M_{z^{\prime }}=\frac{\mathbf{m}}{2}(\frac{g^{\prime }}{g_{0}})\sin (\theta
_{s}).
\end{equation}

From eq (\ref{h}) the free Lagrangian associated with the field $\Xi _{2}$
is given by: 
\begin{eqnarray}
\frak{L}_{free}^{\Xi _{2}} &=&\frac{1}{8g_{0}^{2}}\Upsilon ^{2}(\left| \Xi
_{2}+1\right| ^{2}-1)^{2} \\
&=&\frac{1}{2}\left( \frac{\bar{g}}{g_{0}}\right) ^{2}\left( \frac{\Upsilon 
}{\mathbf{m}}\right) ^{2}\phi _{2}^{2}+\text{higher order terms}  \nonumber
\end{eqnarray}
where 
\begin{equation}
\Upsilon ^{2}=tr(\tilde{M}_{\{ud\}}^{2}+\frac{1}{3}\tilde{M}_{\{\nu
e\}}^{2}).
\end{equation}

Hence 
\begin{equation}
m_{\phi _{2}}=(\frac{\bar{g}}{g_{0}})\left( \frac{\Upsilon }{\mathbf{m}}%
\right) .
\end{equation}

If we take $\bar{g}=g_{0},$ then 
\begin{equation}
m_{\phi _{2}}=\sqrt{\frac{tr(\tilde{M}_{\{ud\}}^{2}+\frac{1}{3}\tilde{M}%
_{\{\nu e\}}^{2})}{tr(M_{u}M_{u}^{*}+M_{d}M_{d}^{*}+\frac{1}{3}%
M_{e}M_{e}^{*}+\frac{1}{3}M_{\nu }M_{\nu }^{*})}}.
\end{equation}

Neglecting the mass of fermions in respect to the mass of the top quark
gives:

\begin{equation}
m_{\phi _{2}}\approx \sqrt{\frac{tr(\tilde{M}_{\{ud\}}^{2})}{%
tr(M_{u}M_{u}^{*})}=}\frac{3}{2}m_{t}
\end{equation}
which is the same relation found by Wulkenhaar for the neutrinos mass matrix 
$M_{\nu }\neq 0$ \cite{Wulkenhaar1}.

Let us now turn to the mixing angles, eq (\ref{free}) looks like its
counterpart in classical model provided that: 
\begin{equation}
g_{s}=g_{L}=g_{R}=g_{0}
\end{equation}
and 
\begin{equation}
g^{\prime }=\sqrt{\frac{3}{2}}g_{0}  \label{unification}
\end{equation}
which give: 
\begin{equation}
\sin ^{2}(\theta _{W})=\frac{3}{8}
\end{equation}
and 
\begin{equation}
\sin ^{2}(\theta _{S})=\frac{3}{5}
\end{equation}
which are the same angles given by $SO(10)$ GUT. However du to eq.(\ref
{unification}) the grand unification for coupling constants is not achieved.

\section{Conclusion}

We reformulate the left-right model in the framework of nonassociative
geometry. We have found that the left-right symmetric model does not exhibit
the parity violation, however NAG gives the mixing angles predicted by
SO(10) GUT and at the tree level we have the mass relations

\begin{equation}
M_{W_{L}}=\frac{1}{2}(\frac{g_{L}}{g_{0}})m_{t}
\end{equation}

\begin{equation}
M_{W_{R}}=\frac{1}{2}(\frac{g_{R}}{g_{0}})m_{t}
\end{equation}

\begin{equation}
M_{Z}=\frac{M_{W_{L}}}{\cos (\theta _{W})}
\end{equation}

\begin{equation}
M_{Z^{\prime }}=\frac{g^{\prime }\sin (\theta _{S})}{g_{R}}M_{W_{R}}
\end{equation}

and

\begin{equation}
M_{H}=\frac{3}{2}m_{t}
\end{equation}

where $m_{t}$ is the top mass.

The unification of the coupling constants is not achieved using this
reformulation.

The main result of this work is that nonassociative geometry, like
noncommutative geometry, does not provide a geometrical explanation of the
parity violation.

\textbf{Acknowledgments}

One of us (A. B.) would like to express his sincere thanks to Dr. R.
Wulkenhaar and Dr. F.Cuypers for private communications.

This work was supported by the Algerian Minist\`{e}re de l'Enseignement
Sup\'{e}rieur et La Recherche Scientifique under the contract D2501/01/05/97.

\end{document}